\documentclass[letter]{IEEEtran}
\usepackage{algorithm,algpseudocode,graphicx}  
\usepackage{amsfonts}
\usepackage{tabularx}
\usepackage{subfigure}
\usepackage{amssymb}
\usepackage{booktabs} 
\usepackage{multirow} 
\usepackage{epsfig}
\usepackage{epstopdf} 

\usepackage[noadjust]{cite}
\newtheorem{theorem}{Theorem}
\newtheorem{remark}{Remark}

\newtheorem{proposition}{Proposition}

\usepackage[cmex10]{amsmath} 
\allowdisplaybreaks  
   
 \makeatletter 
 \def\@eqnnum{{\normalsize \normalcolor (\theequation)}} 
  \makeatother
\begin{document} 
\title{Jointly Optimal Spatial Channel Assignment\\ and Power Allocation for MIMO SWIPT Systems}  
\author{Deepak Mishra,~\IEEEmembership{Member,~IEEE} and George~C.~Alexandropoulos,~\IEEEmembership{Senior~Member,~IEEE}
\thanks{D. Mishra was with the Department of Electrical Engineering, IIT Delhi, 110016 New Delhi, India. He is now  with the Department of Electrical Engineering (ISY) at Link\"oping University, 581 83 Link\"oping, Sweden (e-mail: deepak.mishra@liu.se).}
\thanks{G. C. Alexandropoulos is with the Mathematical and Algorithmic Sciences Lab, Paris Research Center, Huawei Technologies France SASU, 92100 Boulogne-Billancourt, France. The views expressed here are his own and do 	not represent Huawei's ones. (e-mail: george.alexandropoulos@huawei.com).}
}
\maketitle

\begin{abstract}
The joint design of spatial channel assignment and power allocation in {Multiple Input Multiple Output (MIMO)} systems capable of Simultaneous Wireless Information and Power Transfer (SWIPT) is studied. Assuming availability of channel state information at both communications ends, we maximize the harvested energy at the multi-antenna receiver, while satisfying a minimum information rate requirement for the MIMO link. We first derive the globally optimal eigenchannel assignment and power allocation design, and then present a practically motivated tight closed-form approximation for the optimal design parameters. Selected numerical results verify the validity of the optimal solution and provide useful insights on the proposed designs as well as the pareto-optimal rate-energy tradeoff. 
\end{abstract}  
\IEEEpeerreviewmaketitle 

\section{Introduction}\label{sec:introduction} 
Energy sustainability of fifth generation (5G) wireless networks has become a major design challenge due to the raising power consumption demands. Wireless {Energy Harvesting (EH)} has the potential to combat this problem paving the way for the {Simultaneous Wireless Information and Power Transfer (SWIPT)} concept~\cite{SWIPT_modern}. {This concept has been lately investigated for {Multiple Input Multiple Output (MIMO)} systems~\cite{MIMO_SWIPT,JPDT-MU-MIMO-R2,MU-MIMO-SWIPT-R1,Spatial,Spatial2} aiming at exploiting their spatial dimension for efficiently handling the fundamental SWIPT rate-energy tradeoff~\cite{SWIPT_modern}.}  

{As existing {Radio Frequency (RF)} EH circuits are unable to directly perform {Information Decoding (ID)} \cite{MIMO_SWIPT, SWIPT_modern}, there is a need for practical Receiver (RX) architectures profiting from SWIPT. {Time Switching (TS)} and {Power Splitting (PS)}~\cite{MIMO_SWIPT} were the first proposed RX architectures separating the received signal either in time or in power domain for carrying out both EH and ID. Efficient schemes for  optimizing transmit beamforming and receive PS in multiuser MIMO systems have been presented to jointly optimize EH and ID performance for a given TX power budget constraint~\cite{JPDT-MU-MIMO-R2}, or vice-versa~\cite{MU-MIMO-SWIPT-R1}.  Very recently, {Spatial Switching (SS)} RX architecture was proposed in~\cite{Spatial}, which assumes availability of {Channel State Information (CSI)} at both the multiantenna {Transmitter (TX)} and RX, as conventional closed-loop MIMO techniques do. Using the knowledge of the channel matrix for effectively performing its {Singular Value Decomposition (SVD)} with adequate TX-RX signal processing, the received power over some spatial eigenchannels can be used for EH and over the remaining ones for ID. Late advances in analog signal processing designs include an eigenmode transceiver technique \cite{Analog2} that can perform SVD of the channel matrix in the analog domain without impacting the energy content of the received signal. Such analog designs along with the emerging hybrid beamforming techniques \cite{Hybrid-Survey} can enable the practical implementation of SS-based SWIPT.}   

{Although TS and PS architectures have been largely investigated, SS operation for optimized utilization of the available spatial degrees of freedom in MIMO channels is still in its infancy~\cite{Spatial,Spatial2}}. In~\cite{Spatial}, the joint eigenchannel assignment and {Power Allocation (PA)} problem was investigated for minimizing the total transmit power required to meet rate and energy requirements. Joint antenna selection and SS for maximizing the energy efficiency of MIMO SWIPT systems was studied in~\cite{Spatial2}. {However, these suboptimal designs~\cite{Spatial,Spatial2} are based on convex relaxations and rely on iterative optimization techniques. Motivated by the fact that the performance of practical SWIPT systems is bottlenecked by very low RF energy transfer efficiency~\cite{ComMag}, we study in this paper the joint eigenchannel assignment and PA in MIMO SWIPT systems with SS reception for maximizing the harvested {Direct Current (DC)} power, while meeting a minimum rate requirement.} {We obtain the globally {Optimal Eigenchannel Assignment (OEA)} and PA design for the case where perfect CSI is available at both TX and RX. We also present a closed-form tight approximation for the jointly optimized design parameters. Selected results show the impact of various system parameters on the optimized SS-based MIMO SWIPT performance and how this performance compares with conventional PS reception.}

\emph{Notations}: Vectors and matrices are denoted by boldface lowercase and capital letters, respectively. The transpose and Hermitian transpose of $\mathbf{A}$ are denoted by $\mathbf{A}^{\rm T}$ and $\mathbf{A}^{\rm H}$, respectively, and its trace by $\mathrm{tr}\left(\mathbf{A}\right)$. $[\mathbf{a}]_i$ stands for $\mathbf{a}$'s $i$-th element and ${\rm diag}(\mathbf{a})$ denotes a square diagonal matrix with $\mathbf{a}$ placed in its main diagonal. $\mathbb{C}$ is the complex number set.

\section{System Model and Problem Formulation}\label{sec:problem}  
\subsection{System Model}\label{sec:system_model}
Consider a MIMO SWIPT system comprising of a TX equipped with $N_T$ antennas and a RX having $N_R$ antennas. We assume that RX can be powered by EH from the energy-rich TX which acts as an integrated RF energy supply and information source. We consider a frequency flat MIMO fading channel $\mathbf{H}\in\mathbb{C}^{N_R\times N_T}$ that remains constant during one transmission time slot and changes independently from one slot to the next. The entries of $\mathbf{H}$ are assumed to be independent {Zero Mean Circularly Symmetric Complex Gaussian (ZMCSCG)} random variables with variance $\sigma_h^2$ depending on propagation losses; this assumption ensures that the rank of $\mathbf{H}$ is $r=\min\{N_R,N_T\}$. The discrete-time baseband received signal $\mathbf{y}\in\mathbb{C}^{N_R\times 1}$ at RX can be mathematically expressed as 
\begin{equation}\label{eq:sys_model}
\mathbf{y} \triangleq \mathbf{H}\mathbf{x} + \mathbf{n},
\end{equation}
where $\mathbf{x}\in\mathbb{C}^{N_T\times 1}$ denotes the transmitted signal with covariance matrix $\mathbf{S}\triangleq\mathbb{E}\{\mathbf{x}\mathbf{x}^{\rm H}\}$ and $\mathbf{n}\in\mathbb{C}^{N_R\times 1}$ represents the {Additive White Gaussian Noise (AWGN)} vector having ZMCSCG statistically independent entries each with variance $\sigma_n^2$. We also make the usual assumption that the signal elements are statistically independent with the noise elements. For the transmitted signal we assume that there is an average power constraint across all TX antennas denoted by $\mathrm{tr}\left(\mathbf{S}\right)\le P_T$.

Assuming the availability of perfect CSI at both TX and RX, we consider SS reception~\cite{Spatial} according to which RX chooses some eigenchannels for EH and some for ID. To accomplish SS reception, TX precodes the information signal $\mathbf{s}\in\mathbb{C}^{r\times 1}$ with the unitary $\mathbf{V}\in\mathbb{C}^{N_T\times r}$ (i$.$e$.$, the transmitted signal is given by $\mathbf{x}\triangleq\mathbf{V}\mathbf{s}$) and RX combines the elements of $\mathbf{y}$ with the unitary $\mathbf{U}^{\rm H}\in\mathbb{C}^{r\times N_R}$. The latter matrices are obtained from the reduced SVD of $\mathbf{H}$, i$.$e$.$, $\mathbf{H}=\mathbf{U}\boldsymbol{\Lambda}\mathbf{V}^{\rm H}$, where $\boldsymbol{\Lambda}\triangleq{\rm diag}([\lambda_1\,\lambda_2\,\ldots\,\lambda_r])\in\mathbb{C}^{r\times r}$ contains the $r$ singular values of $\mathbf{H}$. With the latter processing, the MIMO link of \eqref{eq:sys_model} can be decomposed into the following $r$ parallel SISO eigenchannels 
\begin{equation}\label{eq:rxS_e}
[\mathbf{U}^{\rm H}\mathbf{y}]_k = \lambda_k[\mathbf{s}]_k+ [\mathbf{U}^{\rm H}\mathbf{n}]_k,\quad k=1,2,\ldots,r.
\end{equation}
For each $k$-th eigenchannel we associate the binary variable $\rho_k$. When $\rho_k=1$, the $k$-th eigenchannel is dedicated for EH, while for $\rho_k=0$ it is used for ID. Hence, it follows from \eqref{eq:rxS_e} that the achievable rate of the considered system is given by
\begin{equation}\label{eq:R}
R \triangleq  \textstyle\sum_{k=1}^{r}\log_2\left(1+\sigma_n^{-2}{\left(1-\rho_k\right)p_k|\lambda_k|^2}\right),
\end{equation} 
where $p_k\triangleq\mathbb{E}\{|[\mathbf{s}]_k|^2\}$. Using the unit channel block duration assumption, the total harvested energy (or power) is given by
\begin{equation}\label{eq:Eh}
P_H \triangleq \textstyle\sum_{k=1}^{r} P_{H,k}=\sum_{k=1}^{r} \eta(P_{R,k})\,P_{R,k},
\end{equation}
where $P_{R,k} \triangleq \rho_kp_k|\lambda_k|^2$and $P_{H,k}$ denote the received RF power for EH and the harvested DC power over the $k$-th eigenchannel, respectively. Function $\eta(\cdot)$ represents the RF-to-DC rectification efficiency, which is in general a \textit{nonlinear} positive function of the received RF power $P_{R,k}$ for EH~\cite{CAMSAP17}. Despite this nonlinear relationship, we note that $P_{H,k}$ is monotonically nondecreasing in $P_{R,k}$ for any practical RF EH circuit~\cite{CAMSAP17,Ref1} due to the law of energy conservation.

 \subsection{Problem Formulation}\label{sec:opt}
In this letter, we are interested in the joint optimal assignment and PA of the available eigenchannels for maximizing $P_H$, while satisfying an underlying minimum rate requirement $\bar{R}$. Using \eqref{eq:R} and \eqref{eq:Eh} we formulate the following optimization problem for the joint design of $\rho_k$ and $p_k$ $\forall$ $k=1,2,\ldots,r$:
\begin{eqnarray*}\label{eqOPT1}
\begin{aligned} 
\mathcal{OP}:\;&\underset{\{\rho_k\}_{k=1}^r,\{p_k\}_{k=1}^r}{\text{max}}\textstyle\sum_{k=1}^{r} \eta(\rho_kp_k|\lambda_k|^2)\rho_kp_k|\lambda_k|^2,\quad\text{s.t.:}\\
&({\rm C1}): \textstyle\sum_{k=1}^{r}\log_2\left(1+\sigma_n^{-2}\left(1-\rho_k\right)p_k|\lambda_k|^2\right)\ge \bar{R},\\
&({\rm C2}): \textstyle\sum_{k=1}^{r}p_k\le P_T,\,({\rm C3}): p_k\ge0\,\,\forall k=1,2,\ldots,r,\\
&({\rm C4}): \rho_k\in\{0,1\}\,\,\forall k=1,2,\ldots,r.
\end{aligned}\hspace{-3mm}
\end{eqnarray*}
Constraints $({\rm C1})$ and $({\rm C2})$ refer respectively to the minimum rate and maximum TX power requirements, whereas $({\rm C3})$ and $({\rm C4})$ include the boundary conditions for $p_k$'s and $\rho_k$'s, respectively. {We next present the $\mathcal{OP}$'s infeasibility condition.}
\begin{remark}\label{rem:feas}
{$\mathcal{OP}$ is not feasible when there is no RF power left for EH, i$.$e$.$, $p_r=0,$ after meeting the rate requirement $\bar{R}\hspace{-0.5mm}\ge\hspace{-0.5mm}R_{\max}\hspace{-0.5mm}\triangleq\hspace{-0.5mm}\textstyle\sum_{j=1}^{r-1}\hspace{-0.5mm}\log_2\hspace{-0.5mm}\left(1+\sigma_n^{-2}\bar{p}_j|\lambda_j|^2\right)$ using the best gain $r-1$ eigenchannels for ID. Here, PA $\bar{p}_j$ $\forall$ $j=1,2,\ldots,r-1$ is obtained from the standard waterfilling approach~\cite{WFG} as
\begin{eqnarray}\label{eq:PAWF}
\bar{p}_j=\begin{cases}
\hat{p}_{\omega}+\sigma_n^2\left(|\lambda_{\omega}|^{-2}-|\lambda_j|^{-2}\right), & \text{$j=1,2,\ldots,\omega$}\\
0, & \text{$\omega< j\le r-1$}.
\end{cases}
\end{eqnarray}}  
In \eqref{eq:PAWF}, $\hat{p}_{\omega}\triangleq\frac{1}{\omega}\big({P_T-\sigma_n^2\textstyle\sum_{j=1}^{\omega-1}\left(|\lambda_{\omega}|^{-2}-|\lambda_j|^{-2}\right)}\big)$ and $\omega\triangleq \max\left\lbrace j\,\Big| P_T-\sigma_n^2\textstyle\sum_{i=1}^{j-1}\left(|\lambda_j|^{-2}-|\lambda_i|^{-2}\right)  >0,\, j \in \mathcal{E}\!\setminus \{r\} \right\rbrace\!.$
\end{remark}
 
$\mathcal{OP}$ is a nonlinear nonconvex combinatorial optimization problem including the nonlinear function $\eta(\cdot)$ and the binary variables $\rho_k$'s in both the objective and constraints. We next present in Propositions~\ref{porp:eigEH} and~\ref{prop:Eqv} the two key properties of $\mathcal{OP}$ that will be exploited to obtain its globally optimal solution.
\begin{proposition}\label{porp:eigEH}
Assigning only one eigenchannel for EH in the optimization problem $\mathcal{OP}$ is optimal.  
\end{proposition}
\begin{IEEEproof}
The proposition is proved by contradiction. Suppose that eigenchannels $e,\upsilon\in\mathcal{E}\triangleq\{1,2,\ldots,r\}$ with $e\neq\upsilon$ and $r>2$ having respective gains $\lambda_e$ and $\lambda_\upsilon$, where $|\lambda_e|^2>|\lambda_\upsilon|^2$, are assigned for EH, and $P_{T,H}$ denotes the transmit power available for EH. We assume that the remaining power $P_T-P_{T,H}$ is allocated to the remaining $r-2$ eigenchannels for ID to meet $\bar{R}$. If $p_e$ is the power allocated to $e$, then $p_\upsilon\triangleq P_{T,H}-p_e$ represents the power given to $\upsilon$. With the above, the total harvested DC power $P_H $ at RX is given by
\begin{align}\label{eq:prop1}
P_H = & \,\eta(p_e|\lambda_e|^2)\,p_e|\lambda_e|^2 + \eta(p_\upsilon|\lambda_\upsilon|^2)\,p_\upsilon|\lambda_\upsilon|^2\nonumber\\
\stackrel{(a)}{<}& \,\eta(P_{T,H}|\lambda_e|^2)\,P_{T,H}|\lambda_e|^2,
\end{align}
where $(a)$ follows from the nondecreasing nature of $P_{H,e}$ and $P_{H,\upsilon}$ over $P_{R,e}$ and $P_{R,\upsilon}$, respectively, along with the assumed gain ordering and $P_{T,H}\triangleq p_e+p_\upsilon$. This proves that allocating $P_{T,H}$ to only $e$ (representing best gain eigenchannel available for EH) always results in the highest harvested power $P_H$. 
\end{IEEEproof} 
 
\begin{proposition}\label{prop:Eqv}
Maximizing the harvested DC power $P_H=\eta(p_e|\lambda_e|^2)\,p_e|\lambda_e|^2$ over the $e$-th eigenchannel, while meeting ID rate $\bar{R}$ over the eigenchannel set $\mathcal{E}_e\triangleq\mathcal{E}\setminus\{e\}$, is equivalent to maximizing the received RF power $P_R\triangleq p_e|\lambda_e|^2$.
\end{proposition} 
\begin{IEEEproof}
As $P_H$ is a nondecreasing positive function of $P_R$ \cite{CAMSAP17,Ref1}, maximizing $P_H$ and $P_R$ are equivalent~\cite{Baz}. As a result both problems share the same solution set.
\end{IEEEproof}

\section{Optimal Spatial Resource Allocation}\label{sec:global} 
{Combining Remark~\ref{rem:feas} and Propositions~\ref{porp:eigEH} and~\ref{prop:Eqv} for $\bar{R}\le R_{\max}$, $\mathcal{OP}$'s optimal solution is obtained in two steps. We first solve iteratively the following problem $\mathcal{OP}1$ for each one eigenchannel $e\in\mathcal{E}$ dedicated to EH, while the remaining $r-1$ ones are intended for ID. We then select the assignment $e$ for EH among the $r$ possible values that yields maximum $P_R$.}
\begin{eqnarray*}\label{eqOPT2}
\begin{aligned} 
\mathcal{OP}1:\;&\underset{p_e\,\textrm{with}\,e\in\mathcal{E},\,p_j\,\textrm{with}\,j\in\mathcal{E}_e}{\text{max}}p_e|\lambda_e|^2,\quad\text{s.t.:}\,\,({\rm C2}),\,({\rm C3}),\\
&({\rm C5}): \textstyle\sum_{j\in\mathcal{E}_e}\log_2\left(1+\sigma_n^{-2}p_j|\lambda_j|^2\right)\ge \bar{R}.
\end{aligned}\hspace{-3mm}
\end{eqnarray*}
{Since $\mathcal{OP}1$ is a convex problem having a linear objective, real variables $p_j$ $\forall j\in\mathcal{E},$ and convex constraints, the globally optimal solution of $\mathcal{OP}$ can be derived as shown next.}
\begin{theorem}\label{th:GWF}
OEA $e^*$ for EH that maximizes $P_R$ (or $P_H$) is
\begin{equation}\label{eq:GWF0}
e^* =  \underset{1\le e\le r}{\operatornamewithlimits{argmax}}\left\lbrace|\lambda_e|^2\left(P_T-\textstyle\sum_{j\in\mathcal{E}_e}p_j^*\right)\right\rbrace
\end{equation} 
with assigned power $p_{e^*} = P_T-\textstyle\sum_{j\in\mathcal{E}_{e^*}}p_j^*$. The OPA $p_j^*$ $\forall$ $j\in\mathcal{E}_e$ for the remaining $r-1$ eigenchannels for ID is
\begin{eqnarray}\label{eq:GWF1}
p_j^* = \begin{cases}
p_{s}^*+\sigma_n^2\left(|\lambda_s|^{-2}-|\lambda_j|^{-2}\right),   & \text{$j\le s\,\,\textrm{and}\,\,j\neq e$}\\
0, & \text{$s<j\le r\,\,\textrm{and}\,\,j\neq e$}.
\end{cases}\hspace{-3mm} 
\end{eqnarray}  
In the latter expression, the water level step $s$ is defined as
\begin{eqnarray}\label{eq:GWF2}
s\triangleq \max \left\lbrace k\,\mathrel{}\middle|\mathrel{} 2^{\bar{R}}>\textstyle\prod_{j=1,j\neq e}^{k}|\lambda_j|^2|\lambda_{k}|^{-2}\,\,\textrm{and}\,\,k\in\mathcal{E}_e\right\rbrace
\end{eqnarray}
and the power level $p_{s}^*$ for this step is given by
\begin{equation}\label{eq:GWF3}
p_{s}^* = \sigma_n^2\Big({2^{\frac{\bar{R}}{\kappa-1}}}{\big(\textstyle\prod_{j=1,j\neq e}^{s}|\lambda_j|^2\big)^{\frac{1}{1-\kappa}}}-|\lambda_s|^{-2}\Big),
\end{equation}
where $\kappa$ denotes the number of eigenchannels with nonzero OPA $p_k^*$ $\forall k\in\mathcal{E}$, while holding $p_e^*>0$.
\end{theorem} 
\begin{IEEEproof}
As shown in \eqref{eq:GWF0}, the OEA $e^*$ for EH is the one that results in the maximum product of channel gain and power available for EH, while meeting $\bar{R}$. {Clearly, one may choose between assigning an eigenchannel with large gain for EH leaving the weaker ones for ID or allocating high power for EH using an eigenchannel with weak gain, while the large gain eigenchannels are devoted for ID. This tradeoff needs to be optimally solved.} To find OPA in $\mathcal{OP}1$ for a given eigenchannel assignment $e$ for EH, we next formulate an equivalent optimization problem $\mathcal{OP}2$ that finds the minimum power required over $\mathcal{E}_e$ for ID to satisfy $\bar{R}$:
\begin{eqnarray*}\label{eqOPT3} 
\mathcal{OP}2:\;\underset{p_j\,\textrm{with}\,j\in\mathcal{E}_e}{\text{min}}\textstyle\sum_{j\in\mathcal{E}_e}p_j,\quad\text{s.t.:}\,\,({\rm C3}),\,({\rm C5}).\hspace{-3mm}
\end{eqnarray*}
Since the objective in $\mathcal{OP}2$ is linear and the constraints are convex, its globally optimal solution can be obtained from its Karush-Kuhn-Tucker (KKT) point~\cite{Baz}. Associating the Lagrange multipliers $\mu_j$ $\forall j\in\mathcal{E}_e$ with constraint $({\rm C3})$ and $\nu$ with $({\rm C5})$, the Lagrangian function of $\mathcal{OP}2$ is given by  
\begin{eqnarray}\label{eq:lang}
L\left(p_j\right) \!=\!  \textstyle\sum\limits_{j\in\mathcal{E}_e}\!p_j(1-\mu_j) -\nu\bigg(\!\textstyle\sum\limits_{j\in\mathcal{E}_e}\!\log_2\!\left(\!1+\frac{p_j|\lambda_j|^2}{\sigma_n^2}\!\right)\!-\!\bar{R}\!\bigg).\!\!
\end{eqnarray} 
The corresponding KKT conditions can be obtained from $({\rm C3})$, $({\rm C5})$, $\mu_j,\nu\ge0$, $p_j\mu_j=0$ $\forall j\in\mathcal{E}_e$, along with  
\begin{subequations}\label{eq:KKT12}
\begin{equation}\label{eq:KKT1}
\frac{\partial L}{\partial p_j}=1{-\mu_j}-\frac{\nu|\lambda_j|^2}{\ln 2 \left(\sigma_n^2+p_j|\lambda_j|^2\right)}=0,
\end{equation}
\begin{equation}\label{eq:KKT2}
\nu\,\big[\,\textstyle\sum_{j\in\mathcal{E}_e}\log_2\left(1+\sigma_n^{-2}{p_j|\lambda_j|^2}\right)-\bar{R}\,\big]=0.
\end{equation}
\end{subequations}
By realizing that $({\rm C5})$ is always satisfied at strict equality when there exists sufficient PA over $\mathcal{E}_e$ meeting $\bar{R}$, we observe from \eqref{eq:KKT12} that the optimal $\nu$, denoted by $\nu^*$, is strictly positive. Using $\tilde{\nu}^*\triangleq\frac{\nu^*}{\ln 2}$, the OPA for each $j$-th eigenchannel used for information communication is given by  
\begin{equation}\label{eq:WF}
p_j^*=\tilde{\nu}^*\left(1-\mu_j\right)^{-1}-{\sigma_n^2}|\lambda_j|^{-2},\quad\forall j\in\mathcal{E}_e.
\end{equation} 
With $s$ representing the largest index with positive PA for ID, i$.$e$.$, $p_i=0$ $\forall$ $i>s$, \eqref{eq:WF} can be rewritten as
\begin{equation}\label{eq:WF1}
\textstyle p_j^*=\left(\frac{1-\mu_s^*}{1-\mu_j^*}\right)\left(p_s^*+\frac{\sigma_n^2}{|\lambda_s|^2}\right)-\frac{\sigma_n^2}{|\lambda_j|^2},\quad\forall j,s\in\mathcal{E}_e,
\end{equation} 
where $\mu_j^*=0$ $\forall\,j\le s$ and $\mu_j^*=1-\left(p_s^*+{\sigma_n^2}|\lambda_s|^{-2}\right)\sigma_n^{-2}$ $\times|\lambda_j|^{2}$ $\forall$ $j>s$. Using the definition of $s$, the maximum number of eigenchannels with nonzero PA is represented by $\kappa$ for $p_e>0$. With $\nu^*>0$, substituting \eqref{eq:WF1} into \eqref{eq:KKT2} yields
\begin{eqnarray}\label{eq:WF2}
\left(1+\sigma_n^{-2}{p_s^*\,|\lambda_s|^2}\right)^{\kappa-1}=2^{\bar{R}}\left({\textstyle\prod_{j=1,j\neq e}^{s}|\lambda_j|^2|\lambda_s|^{-2}}\right)^{-1},
\end{eqnarray}
which solving for $p_s^*$ results in \eqref{eq:GWF3}. Clearly, $p_j^*$ is a strictly decreasing function of the index $j$ since $|\lambda_i|^2\le|\lambda_j|^2$ $\forall$ $i\ge j$, therefore, the value for the water level step $s$ is obtained such that $p_s^*>0$. The latter leads to the condition given by \eqref{eq:GWF2}.
\end{IEEEproof}  

The water level step $s$ refers to the eigenchannel $s\in\mathcal{E}_e$ having the weakest channel gain to which the nonzero power $p_s^*$ is allocated for ID. Each $p_j^*$ for $j<s$ in \eqref{eq:WF1} is thus obtained by adding $p_s^*$ with the difference $\sigma_n^2\left(|\lambda_s|^{-2}-|\lambda_j|^{-2}\right)$ between the level depths of the $s$-th and  $j$-th steps. {Finally, the optimal SS variable $\rho_k$ for $\mathcal{OP}$, as obtained from the OEA $e^*$ for EH in $\mathcal{OP}1$, is given by $\rho_k^*=0$ $\forall$ $k\neq e^*$ with $\rho_{e^*}^*=1$.}

 \subsection{Tight Closed-Form Approximation}\label{sec:APS}
As shown in Theorem~\ref{th:GWF}, our jointly optimal eigenchannel assignment and PA solution includes a waterfilling approach according to which the design parameters are obtained. Now, we capitalize on a key characteristic of available RF EH circuits and present a tight closed-form approximation for $e^*$, $p_{e^*}$, and $p_j^*$ $\forall$ $j\in\mathcal{E}_e$. It is evident from~\cite{Ref1} that the received RF power for EH in SWIPT systems needs to be greater than $-30$dBm {in order for the RF EH circuits to provide nonzero harvested DC power after rectification}. {Since the received noise power spectral density is $-175$dBm/Hz, leading to an average received noise power of around $-100$dBm for SWIPT at $915$ MHz with bandwidths of tens of MHz, and the additional circuits or baseband processing noise is around $-60$dBm~\cite{MIMO_SWIPT,MU-MIMO-SWIPT-R1}, the received {Signal-to-Noise-Ratio (SNR)} in practical SWIPT systems is very high ($>30$dB) so as to meet the relatively high ($\ge-30$dBm) RX energy sensitivity constraint~\cite{Ref1}. This practical high SNR assumption implies ${\sigma_n^2}|\lambda_j|^{-2} \approx  0$, which used in \eqref{eq:WF} gives $p_j^*\approx$ $\tilde{\nu}^*\left(1-\mu_j\right)^{-1}>0$ $\forall j\in\mathcal{E}_e$. Then, $\mu_j=0$ $\forall j\in\mathcal{E}_e$ satisfies the KKT condition $p_j\mu_j=0$ $\forall j\in\mathcal{E}_e$ yielding the asymptotically optimal equal PA $\tilde{p}_j^*=\tilde{\nu}^*>0$ $\forall$$j\in\mathcal{E}_e$. Substituting the latter PA into \eqref{eq:KKT2} with $\nu=\nu^*>0$ gives}
\begin{equation}\label{eq:AP1}
\tilde{p}_j^* = {2^{\frac{\bar{R}}{r-1}}\sigma_n^2}{\left(\textstyle\prod_{i=1,i\neq e}^{r} {|\lambda_i|^2}\right)^{\frac{1}{1-r}}}\quad\forall j\in\mathcal{E}_e,
\end{equation}
which shows that each $\tilde{p}_j^*$ depends on the eigenchannel assignment $e$ for EH. The asymptotically optimal closed-form eigenchannel assignment for EH is therefore given by \begin{equation}\label{eq:AP2}
\tilde{e}^* = {\operatornamewithlimits{argmax}}_{\left(1\le e\le r\right)}\,\left\lbrace|\lambda_e|^2\left(P_T-\left(r-1\right)\tilde{p}_j^*\right)\right\rbrace.
\end{equation} 
Using \eqref{eq:AP2}, $p_j^*$ $\forall$ $j\in\mathcal{E}_{\tilde{e}^*}\triangleq\mathcal{E}\setminus\{\tilde{e}^*\}$ can be computed from \eqref{eq:GWF1} by replacing $e$ with $\tilde{e}^*$. This tight asymptotic approximation $\tilde{e}^*\approx e^*$ avoids the iterative computation of $p_j^*$'s for different $e$ in Theorem~\ref{th:GWF} and thus provides the closed-form jointly optimal solution $\tilde{e}^*$ and $p_j^*$ $\forall$ $j\in\mathcal{E}_{\tilde{e}^*}$ with $p_{\tilde{e}^*}= P_T-\textstyle\sum_{j\in\mathcal{E}_{\tilde{e}^*}}p_j^*$.   
 
\section{Numerical Results}\label{sec:results} 
We investigate the variation of the maximum received RF power $P_R^*$, globally OPA $p_k^*$ $\forall$ $k\in\mathcal{E}$, and globally OEA $e^*$ of our proposed SS design for different system configurations. In the figures that follow we have set $P_T=4$W, $\sigma_n^2=\left\{-100,-70\right\}$dBm, and {$\sigma_h^2=\theta d^{-\alpha}$, where $\theta=0.1$ is the average channel attenuation at unit reference distance, $d=\left\{4,5\right\}$m, and $\alpha=2.5$ is the pathloss exponent.}

\begin{figure}[t]
	\centering 
	{{\includegraphics[width=3.4in]{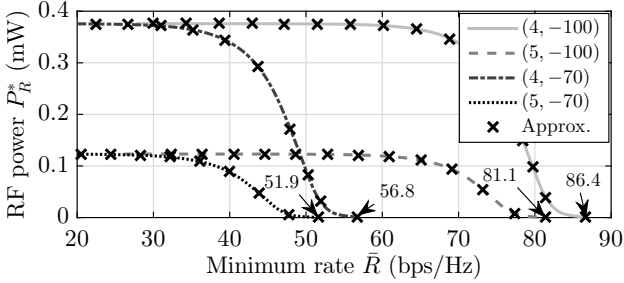} }} 
	\caption{\small {Rate-energy tradeoff of the proposed SS design for $4\times 4$ MIMO systems with different $(d\,\,{\rm in\,\,m},\sigma_n^2\,\,{\rm in\,\,dBm})$ values. $R_{\max}$ for each pair of $d$ and $\sigma_n^2$ values is marked with an arrow.}}
	\label{fig:tradeoff1}
\end{figure}
\begin{figure}[!t]
	\centering  
	\subfigure[Varying $N_R$]
	{{\includegraphics[width=2in]{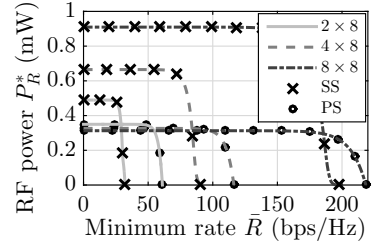} }} 
	\subfigure[Varying $N_T$]
	{{\includegraphics[width=2in]{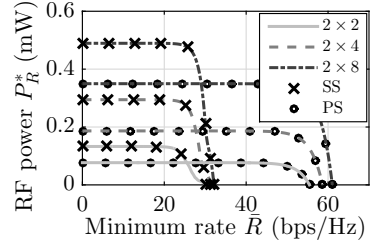} }} 
	\caption{\small {Rate-energy tradeoff of the proposed SS design and PS for different $N_R\times N_T$ MIMO systems with $d=4$m and $\sigma_n^2=-100$dBm.}}
	\label{fig:comp}
\end{figure}

We first plot in Figs$.$~\ref{fig:tradeoff1} and~\ref{fig:comp} the variation of average $P_R^*$ versus different $\bar{R}$ values (also known as rate-energy tradeoff under unit block assumption~\cite{SWIPT_modern,MIMO_SWIPT}) for various MIMO configurations and combinations of $d$ and $\sigma_n^2$. We have used $10^3$ independent channel realizations for the results included in these figures and apart from the jointly optimal solution of $\mathcal{OP}$, the high SNR approximation (``Approx.") is also depicted. As shown in Fig$.$~\ref{fig:tradeoff1} for a $4\times 4$ MIMO system, lower $d$ (i$.$e$.$, lesser propagation losses) and lower $\sigma_n^2$ (i$.$e$.$, less noisy channels) result in improved rate-energy tradeoff. {In addition, as  $e^*=1$ for lower $\bar{R}$ values, the average $P_R^*$ remains almost constant with varying $\bar{R}$ before rapidly falling to $0$ for $\bar{R}$ values approaching the maximum achievable rates $R_{\max}$.} This trend appears also in Fig$.$~\ref{fig:comp} for varying $N_T$ and $N_R$. It is obvious from Fig$.$~\ref{fig:comp}(a) that increasing the rank of the MIMO channel increases the maximum achievable rate and improves the rate-energy tradeoff. However, as depicted in Fig$.$~\ref{fig:comp}(b) for a MIMO channel with a fixed rank, increasing $N_T$ is mainly exploited for providing significant improvement on the average $P_R^*$. It is clear from both figures that the proposed approximation \eqref{eq:AP2} performs sufficiently close to the globally optimal joint design obtained using Theorem~\ref{th:GWF}. {Finally, we sketch in Fig$.$~\ref{fig:comp} the rate-energy tradeoff for the PS architecture. For this case, the PA is given by $\bar{p}_j$ $\forall j\in\mathcal{E}$ as defined in \eqref{eq:PAWF}, but after applying spatial multiplexing \cite{MIMO_SWIPT} over all $r$ eigenchannels available for both EH and ID. With this PA, the ratio $\bar{\rho}$ representing the fraction of received RF power available for EH is obtained by solving $(\mathrm{C}1)$ at equality with $\rho_k=\bar{\rho}$ $\forall k\in\mathcal{E}$. As shown, SS outperforms PS for low values of $\bar{R}$ for all combinations of $N_T$ and $N_R$. However, for $\bar{R}$ values close to $R_{\max}$, PS achieves higher rates than SS. This happens because SS devotes at least one eigenchannel for EH, even for very small targeted $P_H$.} 

\begin{figure}[t]
	\centering 
	{{\includegraphics[width=3.3in]{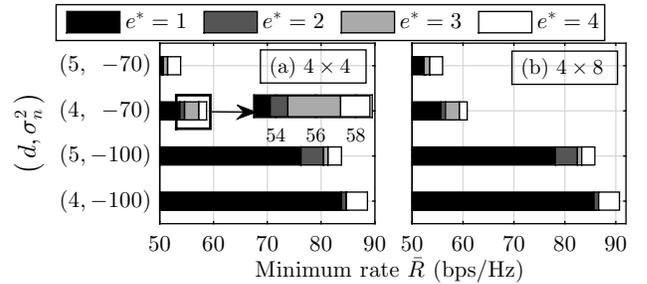} }}
	\caption{\small {OEA $e^*$ for (a) $4\times 4$ and (b) $4\times 8$ MIMO systems with different values for $\bar{R}$ in bps/Hz, $d$ in m, and $\sigma_n^2$ in dBm.}} 
	\label{fig:OEC}
\end{figure}
\begin{figure}[!t]
	\centering 
	{{\includegraphics[width=3.44in]{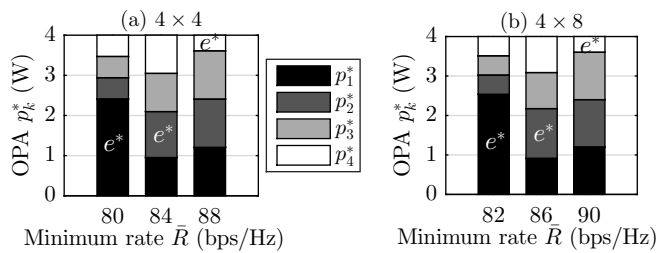} }}
	\caption{\small {OPA $p_k^*$'s and OEA $e^*$ for (a) $4\times 4$ and (b) $4\times 8$ MIMO systems with $d=4$m, $\sigma_n^2=-100$dBm, and different $\bar{R}$ in bps/Hz.}} 
	\label{fig:PA}
\end{figure}

In Figs$.$~\ref{fig:OEC} and~\ref{fig:PA}, we investigate OEA $e^*$ for EH and OPA $p_k^*$ $\forall$ $k\in\mathcal{E}$ with varying $\bar{R}$ as well as combinations of $d$ and $\sigma_n^2$ values for $4\times 4$ and $4\times 8$ MIMO SWIPT systems. For both systems, $r=4$ eigenchannels are assumed to be ordered in decreasing order of their respective gains and the results are plotted for one channel sample. It is observed in Fig$.$~\ref{fig:OEC} that the better the link conditions are and the lower $\bar{R}$ is, the stronger eigenchannel is devoted to EH. {In fact, for most of the feasible rates $\bar{R}$, the best gain eigenchannel $r=1$ is allocated for EH. However, as $\bar{R}$ increases, the weaker eigenchannel is used for EH, and when $\bar{R}$ reaches its maximum value $R_{\max}$, EH is infeasible. Also, increasing $N_T$ leads to improved EH capability at higher $\bar{R}$. As depicted in Fig$.$~\ref{fig:PA}, most of the $P_T$ is allocated to $p_{e^*}$ and $p_j^*\cong\frac{P_T-p_{e^*}}{r-1}$ $\forall$ $j\in\mathcal{E}_{e^*}$ (i$.$e$.$, equal PA) for low $\bar{R}$ values, while $p_{e^*}\rightarrow0$ when $\bar{R}$ approaches $R_{\max}$.}  

\section{Conclusion}\label{sec:conclusion} 
We investigated the joint design of spatial channel assignment and power allocation in SS-based MIMO SWIPT systems. We presented the geometric-waterfilling-based global jointly optimal solution along with a closed-form expression for the asymptotically-optimal eigenchannel assignment. Our numerical investigations, while validating the proposed analysis, provided useful insights on the impact of practical system parameters on the pareto-optimal rate-energy tradeoff.

\bibliographystyle{IEEEtran}  
\bibliography{refs_MIMO_SS_TBF}


\end{document}